\documentclass[twoside,11pt,b5paper]{article}
\usepackage[english]{babel}

\usepackage[]{float,latexsym,times}
\usepackage{amsfonts,amstext,amsmath,amssymb,amsthm}

\usepackage{paralist}
\usepackage{geometry}%
\usepackage{caption2}

\usepackage{titlesec} 

\titleformat{\section}{\large\bfseries}{\thesection.}{.5em}{}
\titlespacing*{\section}{0pt}{*0}{*0}
\titleformat{\subsection}{\normalfont\bfseries}{\thesubsection.}{.5em}{}
\titlespacing*{\subsection}{0pt}{*0}{*0}
\titleformat{\subsubsection}{\normalfont\bfseries}{\thesubsubsection.}{.5em}{}
\titlespacing*{\subsubsection}{0pt}{*0}{*0}

\usepackage{nameref}

\usepackage[%
    colorlinks=true,%
    bookmarksnumbered=true,%
    bookmarksopen=true,%
    citecolor=black,%
    urlcolor=black,%
    unicode=true,           
    breaklinks=true,        
    pdftitle={Optimal Design and Analysis of the Exponentially Weighted Moving Average Chart for Exponential Data},%
    pdfauthor={Aleksey S. Polunchenko, Grigory Sokolov, Alexander G. Tartakovsky},%
]{hyperref}

\usepackage{chngcntr}
\usepackage{enumitem}
\numberwithin{equation}{section} 
\counterwithin{table}{section}
\counterwithin{figure}{section}

\usepackage[authoryear,round]{natbib}

\newcommand{\doi}[1]{\href{http://dx.doi.org/#1}{DOI: #1}}
\newcommand{\nodoi}[1]{}

\usepackage{graphicx}
\graphicspath{{./gfx/}}

\usepackage{bbm}
\usepackage{array}

\usepackage[hang]{subfigure}

\usepackage{url}

\usepackage{multirow}
\usepackage{colortbl}
\usepackage{mathtools} 
\usepackage[all]{hypcap} 
\usepackage{pdfsync}

\newcommand{\ignore}[1]{}

\newcommand{\vect}[1]{\vec{\boldsymbol{#1}}}
\newcommand{\matr}[1]{\boldsymbol{#1}}

\newcommand{\Ind}[1]{\mathbbm{1}_{\{#1\}}}

\theoremstyle{plain} 

\renewcommand{\Pr}{\mathbb{P}} 
\newcommand{\T}{T} 
\newcommand{\Tewma}{T_{\mathrm{EWMA}}} 
\DeclareMathOperator{\EV}{\mathbb{E}} 

\DeclareMathOperator{\LR}{\Lambda}

\DeclareMathOperator{\ADD}{ADD}

\DeclareMathOperator{\ARL}{ARL}
\DeclareMathOperator{\SADD}{SADD}

\DeclareMathOperator{\STADD}{STADD}

\DeclareMathOperator*{\argmin}{arg\,min}

\renewcommand{\le}{\leqslant} 
\renewcommand{\ge}{\geqslant}

\newcommand{\set}[1]{\left\{#1\right\}}

\newcommand{\details}[1]{}

\usepackage{fancyhdr}
\renewcommand{\headrulewidth}{0.0pt}

\begin{document}
\newpage
\setcounter{page}{55}
\fancypagestyle{plain}
{
   \renewcommand{\headrulewidth}{0pt}
   \fancyhf{}
   \fancyhead[L]{\footnotesize Sri Lankan Journal of Applied Statistics Vol(15-2)}
   \fancyfoot[LE,RO]{\thepage}
   \fancyfoot[LO,RE]{\footnotesize IASSL}
   \fancyfoot[CO,CE]{\footnotesize ISSN-1391-4987}
}
\thispagestyle{plain}


\begin{center}
{\Large\textbf{Optimal Design and Analysis of the Exponentially Weighted Moving Average Chart for\\[1.6mm] Exponential Data}}\\
\vspace{0.1cm}
{\large\textbf {Aleksey\ S.\ Polunchenko$^{1}$, Grigory Sokolov$^{1}$ and Alexander\ G.\ Tartakovsky$^{*2}$}}\\
$^{1}$Department of Mathematical Sciences, State University of New York at Binghamton, Binghamton, New York, USA\\
$^{*2}$Department of Statistics, University of Connecticut, Storrs, Connecticut, USA\\
 Corresponding Author: alexander.tartakovsky@uconn.edu\vspace{-0.1in}
\end{center}
\vspace{0.5cm}\
{\footnotesize {Received: date/Accepted: date}}
\vspace{1cm}\
\begin{center}
{\large\textbf {ABSTRACT}}
\end{center}
\vspace{-0.5em}
{ \textit {\noindent We study optimal design of the Exponentially Weighted Moving Average (EWMA) chart by a proper choice of the smoothing factor and the initial value (headstart) of the decision statistic. The particular problem addressed is that of quickest detection of an abrupt change in the parameter of a discrete-time exponential model. Both pre- and post-change parameter values are assumed known, but the change-point is not known. For this change-point detection scenario, we examine the performance of the conventional one-sided EWMA chart with respect to two optimality criteria: Pollak's minimax criterion associated with the maximal conditional expected delay to detection and Shiryaev's multi-cyclic setup associated with the stationary expected delay to detection. Using the integral-equations approach, we derive the exact closed-form formulae for all of the required performance measures. Based on these formulae we find the optimal smoothing factor and headstart by solving the corresponding two bivariate constraint optimization problems. Finally, the performance of the optimized EWMA chart is compared against that of the Shiryaev--Roberts--$r$ procedure in the minimax setting, and against that of the original Shiryaev--Roberts procedure in the multi-cyclic setting. The main conclusion is that the EWMA chart, when fully optimized, turns out to be a very competitive procedure, with performance nearly indistinguishable from that of the known-to-be-best Shiryaev--Roberts--$r$ and Shiryaev--Roberts procedures.}}\\
%
%
\vspace{-0.5em}
%
%
%
%
\section{Introduction}
\label{sec:intro}

Quickest change-point detection is concerned with the design and analysis of procedures for on-line detection of possible changes in the characteristics of an observed random process. Specifically, the process is assumed to be monitored through sequentially made observations (e.g., measurements), and should their behavior suggest the process may have statistically changed, the aim is to conclude so within the fewest observations possible, subject to a tolerable level of the risk of false detection. The area finds applications across many branches of science and engineering: industrial quality and process control (see, e.g.,~\citealp{Ryan:Book2011,Montgomery:Book2012,Kenett+Zacks:Book1998}), biostatistics, economics, seismology, forensics, navigation, cybersecurity (see, e.g.,~\citealp{Tartakovsky+etal:JSM2005,Tartakovsky+etal:SM2006-discussion,Polunchenko+etal:SA2012,Tartakovsy+etal:IEEE-JSTSP2013}), and communication systems---to name a few. A sequential change-point detection procedure is defined as a stopping time adapted to the observed data, at which one stops and declares that apparently a change is in effect.

\noindent The design of a detection procedure comes down to constructing an appropriate detection statistic sensitive to a change, which is compared with a threshold. Once the first exceedance of the threshold occurs, the procedure declares a change. To this end, much of statistical inference in change-point detection uses likelihood ratio (LR), and embraces both the maximum LR argument as well as the Bayesian or quasi-Bayesian theory. For example, Page's~\citeyearpar{Page:B54} Cumulative Sum (CUSUM) ``inspection scheme,'' a popular detection procedure, is based on the maximum LR argument. On the other hand, the Bayesian theory is manifested, e.g., in the Shiryaev~\citeyearpar{Shiryaev:SMD61,Shiryaev:TPA63} procedure, its several limiting cases including the Shiryaev--Roberts (SR) procedure (\citealt{Shiryaev:SMD61,Shiryaev:TPA63} and~\citealt{Roberts:T66}) and the Shiryaev--Roberts--$r$ (SR--$r$) procedure proposed by~\cite{Moustakides+etal:SS11}.

\noindent The focus of this paper is on the Exponentially Weighted Moving Average (EWMA) chart, due to~\cite{Roberts:T59}, who called it the Geometric Moving Average chart. This chart is usually applied to ``raw'' data, i.e., it is not LR-based, and is motivated by considerations of forecasting; see, e.g.,~\cite{Hunter:JQT86}. This not withstanding, the EWMA chart can be also successfully used as a change-point detection procedure. The EWMA statistic is defined as $Z_n=(1-\lambda)Z_{n-1}+\lambda X_n$, $n\ge1$ (i.e., linear in observations), where $Z_0$ and $\lambda\in(0,1]$ are design parameters and $\{X_n\}_{n\ge 1}$ are observations. The initial value $Z_0$ is often referred to as the headstart (see, e.g., \citealt{Lucas+Crosier:T1982}), and $\lambda$ is the smoothing factor. Note that $\{Z_n\}_{n\ge 0}$ is a first-order autoregressive process. It has been shown by~\cite{Lucas+Saccucci:T90} that EWMA can be as powerful as CUSUM for the purposes of detecting a shift in the mean of Gaussian noise. As suggested by~\cite{Hunter:JQT86}, the EWMA chart can be viewed as a compromise between the $\bar{X}$--chart and the CUSUM inspection scheme. Specifically, Hunter observed that, on  one hand, if $\lambda=1$, then the EWMA chart is ``memoryless'' and uses only the most recent observation, $X_n$, ignoring all prior data. This is no different from the $\bar{X}$--chart. On the other hand, if $\lambda$ is close to 0, then the distribution of ``importance weight'' across the past observations is roughly uniform: each observation's influence is about $\lambda$. This ``elephant memory'' is akin to the CUSUM chart, since it is equivalent to repetitively applying the SPRT, which in turn is based on accumulating data assigning equal (namely, unit) weight to each observation.

\noindent While it is apparent that the performance of the EWMA chart is dependent upon the choice of the headstart and the smoothing factor, there seems to be no clear answer as to how does one set each of these characteristics to ``squeeze'' as much as possible out of the chart. The main objective of this work is to begin bridging this gap. The current ``rule of thumb'' is to use small values (such as $0.05\le\lambda\le0.25$)  for small shifts and larger values (between 0.2 and 0.4) for larger shifts. This was demonstrated empirically, e.g., by~\cite{Roberts:T59},~\cite{Lucas+Saccucci:T90},~\cite{Crowder:T87,Crowder:JQT1989,Crowder:JQT1987}, and~\cite{Knoth:SC2005}. Also,~\cite{SrivastavaWu97} considered EWMA to detect a change in the drift of a Brownian motion and obtained asymptotically optimal value for the smoothing factor, but their EWMA procedure is slightly different from the conventional one considered in our paper. A thorough study of the problem of the optimal design of the EWMA chart has also been carried out by~\cite{Gan:JQT1998}.

\noindent More specifically, the bulk of the present paper is centered around the basic minimax change-point detection problem. It assumes the observations are independent with the pre- and post-change distributions fully specified. The particular change-point scenario we focus on is based on exponential data with a surge in the parameter. We employ the conventional one-sided EWMA chart to detect the change. It turns out that for this change-point scenario one can obtain exact closed-form expressions for nearly every performance characteristic of the EWMA chart. This was first accomplished by~\cite{Novikov90}; see also~\cite{Areepong+Sukparungsee:TS2010,Mititelu+etal:ICMA-MU2009,Mititelu+etal:EWJM2010,Sukparungsee+Novikov:JQMA2006,Sukparungsee+Novikov:JQMA2008,Areepong+Novikov:JQMA2008}. We extend Novikov's results using a different approach, exploiting renewal integral equations. Specifically, we derive  exact formulas for the conditional Average Detection Delay (ADD) for an arbitrary value of the change-point and for the Stationary Average Detection Delay (STADD). To the best of our knowledge, this is the first time these formulae are obtained. Our goal is to use these formulae to perform an optimization of EWMA with respect to both the headstart and the smoothing factor. A similar analysis was previously carried out, e.g., by~\cite{Cisar+Cisar:APH2011,Sukparungsee:ApMSci2012,Knoth:StP2005}. However, they did not consider optimizing simultaneously with respect to the headstart and the smoothing factor. Nor did they consider optimizing with respect to the STADD.

\noindent The rest of the paper is organized as follows. In Section~\ref{sec:opt-criteria+det-schemes} we formally state the problem and introduce all of the performance measures and procedures of interest. The exact formulae for EWMA's operating characteristics are derived in Section~\ref{sec:perf_eval+opt}. Using these formulae, in Section~\ref{sec:case-study} we perform a case study with two objectives. First, we apply the obtained formulae to optimize the EWMA chart simultaneously with respect to the smoothing factor and the headstart. This is done for the minimax problem and for the multi-cyclic problem separately. Secondly, once the EWMA chart is fully optimized, its performance is compared against that of the Shiryaev--Roberts procedure in the multi-cyclic setting and against that of the Shiryaev--Roberts--$r$ procedure in the minimax setting. These two procedures are selected to serve as benchmarks for a reason: the Shiryaev--Roberts procedure is exactly optimal in the multi-cyclic sense, and the Shiryaev--Roberts--$r$ is nearly optimal in the minimax sense. The popular CUSUM scheme lacks either of these properties and for this reason is not considered.\\
\vspace{-0.5em}
\section{Optimality Criteria and Detection Procedures}
\label{sec:opt-criteria+det-schemes}

As indicated above, the focus of this work is on two formulations of the change-point detection problem: the minimax formulation and that related to multi-cyclic disorder detection in a stationary regime. In this section, we formulate both problems and introduce the relevant detection procedures studied in the sequel.

\noindent Suppose one can observe a series of independent random observations $\{X_n\}_{n\ge1}$. Statistically, the series is such that for some time index $\nu$, which is referred to as the change-point, $X_1,X_2,\ldots, X_{\nu}$ each posses a known pdf $f(x)$, and $X_{\nu+1},X_{\nu+2},\ldots$ are each drawn from a population with a pdf $g(x)\not\equiv f(x)$, also known. The change-point is the unknown serial number of the last $f(x)$-distributed observation; therefore, if $\nu=\infty$, then the entire series $\{X_n\}_{n\ge1}$ is sampled from distribution with the pdf $f(x)$, and if $\nu=0$, then all observations are $g(x)$-distributed. Our objective is to decide that the change is in effect, raise an alarm and respond. The challenge is to make this decision ``as soon as possible'' past and ``no earlier'' than a certain prescribed limit prior to the true change-point.

\noindent Statistically, the problem is to sequentially differentiate between the hypotheses $\mathcal{H}_k\colon\nu=k\ge0$, i.e., that the change occurs at time moment $\nu=k$, $0\le k<\infty$, and $\mathcal{H}_{\infty}\colon\nu=\infty$, i.e., that the change never takes place. Once the $k$-th observation is made, our decision options are either to accept $\mathcal{H}_k$, and thus declare that the change has occurred, or to reject $\mathcal{H}_k$ and continue observing data.

\noindent To test $\mathcal{H}_k$ against $\mathcal{H}_{\infty}$ one first constructs the corresponding likelihood ratio (LR). Let $\boldsymbol{X}_{1:n}=(X_1,X_2,\ldots,X_n)$ be the vector of the first $n\ge1$ observations. The joint densities of $\boldsymbol{X}_{1:n}$ under $\mathcal{H}_k$ and $\mathcal{H}_{\infty}$ are
\begin{align*}
p(\boldsymbol{X}_{1:n}|\mathcal{H}_{\infty})
&=
\prod_{j=1}^n f(X_j)
\quad\text{and}\quad \\
p(\boldsymbol{X}_{1:n}|\mathcal{H}_k)
&=
\Biggl(\,\prod_{j=1}^{k} f(X_j)\Biggr)\times\Biggl(\,\prod_{j=k+1}^n g(X_j)\Biggr),
\end{align*}
where it is understood that $p(\boldsymbol{X}_{1:n}|\mathcal{H}_{\infty})=p(\boldsymbol{X}_{1:n}|\mathcal{H}_k)$ if $k\ge n$. For the LR, $\LR_{k:n}=p(\boldsymbol{X}_{1:n}|\mathcal{H}_k)/p(\boldsymbol{X}_{1:n}|\mathcal{H}_{\infty})$, one then obtains
\begin{align*}
\LR_{k:n}
&=
\prod_{j=k+1}^n\LR_j,
\;\;\text{where}\;\;
\LR_n
=
\frac{g(X_n)}{f(X_n)},
\end{align*}
and we note that $\LR_{k:n}=1$ if $k\ge n$. We also assume that $\Lambda_0=1$.

\noindent Next, to decide which of the hypotheses $\mathcal{H}_k$ or $\mathcal{H}_{\infty}$ is true, the sequence $\{\LR_{k:n}\}_{1\le k\le n}$ is turned into a detection statistic. To this end, one can either use the maximum likelihood principle and the detection statistic $W_n=\max_{1\le k\le n}\LR_{k:n}$ (maximal LR) or  the generalized Bayesian approach (limit of the Bayesian approach) and the quasi-Bayesian detection statistic $R_n=\sum_{k=1}^n\LR_{k:n}$ (average LR with respect to an improper uniform prior distribution of the change-point). See~\cite{Polunchenko+Tartakovsky:MCAP2012,Polunchenko+etal:JSM2013} for a detailed overview of the different change-point detection approaches.

\noindent Once the detection statistic is chosen, it is supplied to an appropriate sequential detection procedure. To this end, the above two detection statistics -- $\{W_n\}_{n\ge 1}$ and $\{R_n\}_{n\ge 1}$ -- give raise to a myriad of detection procedures. The first detection procedure we will consider is the the Shiryaev--Roberts (SR) procedure, which is defined as the stopping time
\begin{align*}
\mathcal{S}_A
&=
\inf\{n\ge1\colon R_n\ge A\},
\end{align*}
where $A>0$ is a preselected detection threshold, and $R_n=\sum_{k=1}^n\LR_{k:n}$ is the SR statistic, which can also be computed recursively as
\begin{align*}
R_n
&=
(1+R_{n-1})\LR_n,\;\; n\ge1\;\;\text{with}\;\; R_0=0,
\end{align*}
Note that the SR statistic starts from {\em zero}. Hereafter in the definitions of stopping times $\inf\{\varnothing\}=\infty$.

\noindent Another procedure we will also be interested in is the Shiryaev--Roberts--$r$ (SR--$r$) procedure. It is a derivative of the SR procedure proposed by~\cite{Moustakides+etal:SS11} which starts at a fixed but specially designed point $R_0=r$, $0\le r<A$. The stopping time with this new deterministic initialization is defined as
\begin{align*}
\mathcal{S}_A^r
&=
\inf\{n\ge1\colon R_n^r\ge A\},
\end{align*}
where $A>0$ is again a preset detection threshold and the SR--$r$ detection statistic $R_n^r$ is given by the recursion
\begin{align*}
R_n^r
&=
(1+R_{n-1}^r)\LR_n,
\;\; n\ge1~~\text{with}\;\;
R_0^r=r\ge0.
\end{align*}
Note that for $r=0$ this is nothing but the conventional SR procedure.

\noindent Unlike the SR procedure or the SR--$r$ procedure, the EWMA chart proposed by~\cite{Roberts:T59} usually does not use the LR, and is applied to raw data. Certainly this statistic can also be built upon the log-likelihood ratio, replacing $X_n$ in~\eqref{eq:ewma_stat} with $\log [g(X_n)/f(X_n)]$, as it is done in certain publications. Specifically, the EWMA detection statistic is defined as
\begin{align}\label{eq:ewma_stat}
Z_{n}^{\lambda, z} &= (1 - \lambda) Z_{n-1}^{\lambda, z} + \lambda X_{n},\;\;n\ge1~~\text{with}\;\;Z_0^{\lambda, z}=z,
\end{align}
where the smoothing factor $\lambda\in(0,1]$ and the initialization point  $z$ are given. The starting point $Z_0^{\lambda, z}=z$ is usually set to the observations' pre-change mean.
The corresponding stopping can be defined in a number of ways. The most popular way is the two-sided EWMA, which can be effectively used  for detecting changes in both directions. We will be interested in the one-sided EWMA defined as
\begin{align*}
\Tewma ^{\lambda, z}(A)
&=
\inf\{n\ge1\colon Z_n^{\lambda, z} \ge A\},
\end{align*}
where $A>0$ is a preset detection threshold.

\noindent We now proceed with reviewing the optimality criteria whereby one decides which procedure to use. We first set down some additional notation. Let $\Pr_{\nu}(\cdot)$ be the probability measure generated by the observations $\{X_n\}_{n\ge1}$ when the change-point is $\nu\ge0$, and $\EV_{\nu}[\,\cdot\,]$ be the corresponding expectation. Likewise, let $\Pr_{\infty}(\cdot)$ and $\EV_{\infty}[\,\cdot\,]$ denote the same under the no-change scenario, i.e., when $\nu=\infty$.

\noindent Consider first the minimax formulation proposed by~\cite{Pollak:AS85}. The risk of raising a false alarm is measured by the average run length (ARL) to false alarm $\ARL(\T)=\EV_{\infty}[\T]$ and the delay to detection by the maximal average delay to detection $\SADD(\T)=\sup_{0\le\nu<\infty}\ADD_{\nu}(\T)$, where $\ADD_{\nu}(\T)=\EV_{\nu}[\T-\nu|\T>\nu]$ is the conditional average delay to detection. Let $\Delta(\gamma)=\{\T\colon\ARL(\T)\ge\gamma\}$ be the class of detection procedures (stopping times) for which the ARL to false alarm does not fall below a given (desired and {\it a~priori} set) level $\gamma>1$. The problem is to find $\T_{\mathrm{opt}}\in\Delta(\gamma)$ such that
\begin{align}
\SADD(\T_{\mathrm{opt}})
&=
\inf_{\T\in\Delta(\gamma)}\SADD(\T) \quad \text{for every $\gamma>1$}. \label{eq:minimax}
\end{align}
An exact solution to this minimax problem is still an open question. See, e.g.,~\cite{Pollak:AS85},~\cite{Polunchenko+Tartakovsky:AS10},~\cite{Moustakides+etal:SS11}, and~\cite{Tartakovsky+etal:TPA2012} for a related study. However, the SR--$r$ procedure is almost $\SADD(\T)$-optimal. Specifically, as shown by~\cite{Tartakovsky+etal:TPA2012}, the SR--$r$ procedure is asymptotically (as $\gamma\to\infty$) third-order minimax. That is,
\begin{align*}
\SADD(\mathcal{S}_A^r)
&=
\inf_{\T\in\Delta(\gamma)}\SADD(\T)+o(1) ~ \;\text{as}\;\gamma\to\infty,
\end{align*}
where $o(1)\to0$ as $\gamma\to\infty$. Note also that the randomized version of the SR procedure initialized from the quasi-stationary distribution of $R_n$, introduced by \cite{Pollak:AS85}, is also third-order asymptotically optimal. CUSUM chart, on the other hand, is only second-order asymptotically (as $\gamma\to\infty$) optimal. Furthermore,~\cite{Polunchenko+Tartakovsky:AS10} and~\cite{Tartakovsky+Polunchenko:IWAP10} showed the SR--$r$ procedure to be exactly minimax in two particular scenarios. Hence, it is the SR--$r$ procedure (and not the CUSUM chart) that one is to use as a benchmark to compare the EWMA chart against in the minimax setting \eqref{eq:minimax}.

\noindent Note that Pollak's version of the minimax formulation assumes the detection procedure is applied only once; the result is either a false alarm, or a correct (though delayed) detection, and no data sampling is done past the stopping point. This is known as the single-run paradigm. Yet another formulation emerges if one considers applying the same procedure in cycles, e.g., starting anew after every false alarm. This is the multi-cyclic formulation.

\noindent Specifically, the idea is to assume that in exchange for the assurance that the change will be detected with maximal speed, the experimenter agrees to go through a ``storm'' of false alarms along the way. The false alarms are ensued from repeatedly applying the same detection rule, starting from scratch after each false alarm. Put otherwise, suppose the change-point, $\nu$,  is substantially larger than the desired level of the ARL to false alarm, $\gamma>1$. That is, the change occurs in a distant future and it is preceded by a stationary flow of false alarms; the ARL to false alarm in this context is the mean time  between consecutive false alarms. As argued by~\cite{Pollak+Tartakovsky:SS09}, this comes in handy in many surveillance applications, in particular in the area of cybersecurity; see, e.g.,~\cite{Tartakovsky+etal:JSM2005,Tartakovsky+etal:SM2006-discussion,Polunchenko+etal:SA2012,Tartakovsy+etal:IEEE-JSTSP2013}.

\noindent Formally, let $T_1,T_2,\ldots$ denote sequential independent applications of the same stopping time $\T$, and let $\mathcal{T}_{j}=T_{1}+T_{2}+\cdots+T_{j}$ be the time of the $j$-th alarm, $j\ge1$. Let $I_\nu=\min\{j\ge1\colon\mathcal{T}_{j}>\nu\}$ so that $\mathcal{T}_{I_\nu}$ is the point of detection of the true change, which occurs at time instant $\nu$ after $I_\nu-1$ false alarms have been raised. Consider $\STADD(\T)=\lim_{\nu\to\infty}\EV_\nu[\mathcal{T}_{I_\nu}-\nu]$, i.e., the limiting value of the ADD that we will refer to as the {\em stationary ADD} (STADD); then the multi-cyclic optimization problem consists in finding $\T_{\mathrm{opt}}\in\Delta(\gamma)$ such that
\begin{align}
\STADD(\T_{\mathrm{opt}})
&=
\inf_{\T\in\Delta(\gamma)}\STADD(\T) \quad \text{for every $\gamma>1$}. \label{eq:stationary}
\end{align}
For the continuous-time Brownian motion model, this formulation was first proposed by~\cite{Shiryaev:SMD61,Shiryaev:TPA63} who showed that the SR procedure is strictly optimal. For the discrete-time iid model, optimality of the SR procedure in this setting was established by~\cite{Pollak+Tartakovsky:SS09}. Specifically, they showed that if the threshold $A=A_\gamma$ in the SR procedure $\mathcal{S}_{A_\gamma}$ is so selected that $\ARL(\mathcal{S}_{A_\gamma})=\gamma$, then
\[
    \STADD(\mathcal{S}_{A_\gamma})=\inf_{\T\in\Delta(\gamma)}\STADD(\T) \quad \text{for every $\gamma>1$}.
\]
We also note that the CUSUM chart does not possess this optimality property. It therefore makes sense to compare the EWMA chart against the SR procedure with respect to $\STADD(\T)$.\\
\vspace{-0.5em}
\section{EWMA Performance Evaluation and Optimization}
\label{sec:perf_eval+opt}

\noindent Consider a model where the observations are independent and exponentially distributed throughout the entire period of surveillance, and assume that the pre- and post-change mean values are $1$ and $1+\theta$, respectively, where $\theta>0$ (known). This is the exponential iid model we intend to focus on in this work. Formally, the corresponding pre- and post-change densities are
\begin{align*}
f(x)
&=
e^{-x}\Ind{x \ge 0} \quad\text{and}\quad g(x)=\frac{1}{1 + \theta} e^{-x / (1 + \theta)}\Ind{x \ge 0},
\end{align*}
respectively, where $\Ind{\mathcal{A}}$ is an indicator of a set $\mathcal{A}$.

\noindent Recall that our objective is to optimize and compare the performance of the EWMA chart against the performance of the SR procedure in the multi-cyclic setting and against the performance of the SR--$r$ procedure in the minimax setting. To achieve this goal, we need to be able to compute the required performance measures for each of the procedures of interest for the change-point scenario in question. To this end, we will rely on the previous work of~\cite{Moustakides+etal:SS11} and its recent extension due to~\cite{Polunchenko+etal:MIPT2013,Polunchenko+etal:SA2014,Polunchenko+etal:ASMBI2014}, who proposed a generic numerical framework allowing one to compute a large range of performance indices for a broad class of detection procedures with Markovian detection statistics. This condition is clearly fulfilled by all of the procedures of interest. We will therefore proceed to employing the numerical framework by setting up the corresponding integral equations and relations.

\noindent We first set down some notation. Following~\cite{Moustakides+etal:SS11}, let $\mathcal{T}_A^x=\inf\{n\ge1\colon V_n^x\ge A\}$ be a generic detection procedure, whose detection statistic, $\{V_n^x\}_{n\ge0}$, admits the recursion
\[
    V_{n+1}^x=\Psi(V_n^x)+X_n \quad \text{for $n\ge0$}
\]
with $V_0^x=x$, a fixed value (headstart), and $\Psi(x)$ a sufficiently smooth function such that $\Psi(x)\ge0$ for $x\ge0$. Let $\ell(x) =\ARL(\mathcal{T}_A^x)$, $\rho_k(x)=\Pr_\infty(\mathcal{T}_A^x > k)$, $k\ge0$, $\delta_k(x)=\EV_k[(\mathcal{T}_A^x-k)^+]$, $k\ge0$, $\psi(x)=\sum_{k=0}^\infty\EV_k[(\mathcal{T}_A^x-k)^+]=\sum_{k=0}^\infty\delta_k(x)$. From~\cite{Moustakides+etal:SS11}, we have the following integral equations and relations:
\begin{align}
\ell(x)
&=
1+\int_0^A\dfrac{\partial}{\partial y}\Pr_\infty\left(X_1<y-\Psi(x)\right)\ell(y)\,dy, \label{eq:int_arl}\\
\delta_0(x)
&=
1+\int_0^A\dfrac{\partial}{\partial y}\Pr_0\left(X_1<y-\Psi(x)\right)\delta_0(y)\,dy, \label{eq:int_add0}\\
\psi(x)
&=
\delta_0(x)+\int_0^A\dfrac{\partial}{\partial y}\Pr_\infty\left(X_1<y-\Psi(x)\right)\psi(y)\,dy, \label{eq:int_iadd}\\
\rho_{k+1}(x)
&=
\int_0^A\dfrac{\partial}{\partial y}\Pr_\infty\left(X_1<y-\Psi(x)\right)\rho_k(y)\,dy, \;\; \rho_0(x)=1, \label{eq:int_rho}\\
\delta_{k+1}(x)
&=
\int_0^A\dfrac{\partial}{\partial y}\Pr_\infty\left(X_1<y-\Psi(x)\right)\delta_k(y)\,dy. \label{eq:int_delta}
\end{align}

Since $\ADD _k (\mathcal{T} _A ^x) = \EV _k [(\mathcal{T} _A ^x - k)^+] / \Pr _\infty (\mathcal{T} _A ^x > k)$ and
\[
  \STADD (\mathcal{T} _A ^x) = \sum _{k = 0}^\infty \EV _k[(\mathcal{T} _A ^x - k)^+] \left/ \ARL (\mathcal{T}_A^x) \right.,
\]
from the above equations we obtain that
\begin{align}
    \ADD _k(\mathcal{T} _A ^x) = \delta _k(x) / \rho _k(x)
    \quad \text{and} \quad
    \STADD (\mathcal{T} _A ^x) = \psi (x)/ \ell (x). \label{eq:frac:addk_stadd}
\end{align}

\noindent We will now narrow down the above equations to the EWMA chart assuming the exponential iid model introduced at the beginning of this section. To this end, we first remark that the integral equations presented above are Fredholm equations of the second kind. Usually, such equations do not allow for an analytical solution and should be solved numerically. However, as we will show shortly, in the case of EWMA and exponential iid model, one can compute explicitly every single performance measure we are after.

\noindent Our goal is twofold. First, we would like to optimize EWMA over the smoothing parameter $\lambda$ and the headstart $z$ with respect to supremum average detection delay \eqref{eq:minimax} and see how its performance compares to that of the SR--$r$ procedure. Second, we would like to optimize the EWMA procedure with respect to stationary average detection delay \eqref{eq:stationary} and compare it to the SR procedure. As we will see, sloppily chosen $\lambda$ and $z$ may result in considerable performance degradation.

\noindent To optimize performance in the minimax sense, choose
\[
    (\lambda_0 , z_0) = \argmin _{\lambda , z} \SADD (\Tewma ^{\lambda, z}(A))
\]
subject to $\ARL = \gamma$ for a given $\gamma > 1$. To optimize performance in the stationary sense, choose
\[
    (\lambda_0 , z_0) = \argmin _{\lambda , z} \STADD (\Tewma ^{\lambda, z}(A))
\]
subject to $\ARL = \gamma$ for a given $\gamma > 1$.

\noindent Throughout the rest of the paper we will write $\alpha$ for $1 - \lambda $. We will also be using the q-analogues notation, namely a special case of the q-Pochhammer symbol, q-brackets and q-factorials:
  \begin{align*}
      (q; q)_n &= \prod_{j = 1}^n (1 - q^j), \quad (q; q)_0 = 1, \\
      [n]_q ! &= \prod_{j = 1}^n [j]_q = \frac{1 - q}{1 - q} \frac{1 - q^2}{1 - q} \cdots \frac{1 - q^n}{1 - q} = \frac{(q; q)_n}{(1 - q)^n}.
  \end{align*}\\
\vspace{-0.5em}
\subsection{\texorpdfstring{Recovering \boldmath{$\ARL $}}{Recovering ARL}}

    Let $\ell(z)$ denote the average run length to false alarm, $\ARL $, of the EWMA procedure initialized from point $z$. Then \eqref{eq:int_arl} becomes
    \[
        \ell (z) = 1 + \frac{1}{\lambda } e^{\frac{1}{\lambda } \alpha z} \int _{\alpha z} ^A e^{-\frac{1}{\lambda } y} \ell (y) \,dy.
    \]
    Looking for a solution in the form
    \[
        \ell (z) = B_0 + \sum_{n = 1}^\infty B_n \frac{z^n}{n!}
    \]
    one obtains
    \begin{align}
        B_0 = 1 + \frac{1}{\lambda} \sum_{n = 1}^\infty \frac{A^n}{n} \frac{[n - 1]_\alpha !}{(n - 1)!}, \;\;
        B_n = -\frac{1}{\lambda} \alpha ^n [n - 1]_\alpha !, \;\; n \ge 1. \label{eq:arl:coeff}
    \end{align}
    Finally,
    \begin{align}
        \ell(z) &= 1 + \frac{1}{\lambda} \left[\sum_{n = 1}^\infty \frac{(A^n - (\alpha z)^n)}{n} \frac{[n - 1]_\alpha !}{(n - 1)!} \right]\Ind{\alpha z \le A}. \label{eq:arl}
    \end{align}
\vspace{-0.5em}\\
\subsection{\texorpdfstring{Recovering \boldmath{$\SADD$}}{Recovering SADD}}

    Since $\SADD = \sup_{0 \le k < \infty} \ADD _k$, we first start with $\ADD _0$, and \eqref{eq:int_add0} becomes
    \[
        \delta _0 (z) = 1 + \frac{1 }{\lambda (1 + \theta )}e^{\frac{1}{\lambda (1 + \theta )} \alpha z} \int _{\alpha z}^A e^{-\frac{1}{\lambda (1 + \theta )} y} \delta _0 (y) \,dy.
    \]
   Looking for a solution in the form
    \[
        \delta_0(z) = B_0^{(0)} + \sum_{n = 1}^\infty B_n^{(0)} \frac{z^n}{n!},
    \]
    one gets
    \begin{align}
        B_0^{(0)} &= 1 + \frac{1}{\lambda} \sum_{n = 1}^\infty \frac{A^n}{n (1 + \theta )^n} \frac{[n - 1]_\alpha !}{(n - 1)!}, \nonumber \\
        B_n^{(0)} &= -\frac{1}{\lambda} \left( \frac{\alpha }{1 + \theta } \right)^n [n - 1]_\alpha !, \quad n \ge 1, \label{eq:delta_zero:coeff}
    \end{align}
    so
    \begin{align}
        \delta_0(z) &= 1 + \frac{1}{\lambda} \left[\sum_{n = 1}^\infty \frac{(A^n - (\alpha z)^n)}{n (1 + \theta )^n} \frac{[n - 1]_\alpha !}{(n - 1)!} \right] \Ind{\alpha z \le A}. \label{eq:delta_zero}
    \end{align}

  \noindent In order to obtain $\ADD_k$ for $k \ge 1$, recall that $\ADD_k = {\delta_k} / {\rho_k}$ \eqref{eq:frac:addk_stadd}, so we have to compute $\delta_k$ and $\rho _k$. From \eqref{eq:int_rho} and \eqref{eq:int_delta},
    \begin{align*}
        \rho _0(z) = 1, \quad
        \rho _k(z) = \frac{1}{\lambda } e^{\frac{1}{\lambda } \alpha z} \int _{\alpha z} ^A e^{-\frac{1}{\lambda } y} \rho _{k - 1}(y) \,dy, \quad k \ge 1,
    \end{align*}
    and
    \begin{align*}
        \delta _k(z) &= \frac{1}{\lambda } e^{\frac{1}{\lambda } \alpha z} \int _{\alpha z} ^A e^{-\frac{1}{\lambda } y} \delta_{k - 1}(y) \,dy, \quad k \ge 1.
    \end{align*}
    Introduce auxiliary functions
    \[
        h_m (z) = \frac{e^{\frac{1 }{\lambda } \alpha ^{m} z}}{(\alpha ; \alpha )_{m} \, e^{\frac{1 }{\lambda } A}}.
    \]
    Then for $k \ge 1$,
    \begin{align}
        \rho _{k}(z) &\ignore{= 1 - \sum _{n = 1} ^k c_n \, \frac{e^{\frac{1}{\lambda } \alpha ^{k - n + 1} z}}{(\alpha ; \alpha )_{k - n} \, e^{\frac{1}{\lambda } A}}}= 1 - \sum _{n = 1} ^k c_n \, h_{k - n + 1} (z) \, (1 - \alpha ^{k - n + 1}), \label{eq:rho_k}
    \end{align}
    where the constants $\set{c_k}_{k \ge 1}$ can be found from solving the following linear system for $\vect{c} = (c_1, c_2, \cdots , c_k)^\top$:
    \begin{align*}
    \vect{c}
    &=
    (1, 1, \cdots , 1)^\top-\matr{M}\,\vect{c},
    \end{align*}
    where $\matr{M}$ is a lower-triangular $k$-by-$k$ Toeplitz matrix with coefficients $\matr{M}_{i, i} = 0$, and
    \begin{align*}
        \matr{M}_{i + j, j} &= h_i (A) = \frac{e^{\frac{1}{\lambda } \alpha ^{i} A}}{(\alpha ; \alpha )_{i} \, e^{\frac{1}{\lambda } A}} , \quad 1 \le i \le k - 1, \, 1 \le j \le k - i.
    \end{align*}

\begin{figure}[!htb]
        \centering
        \subfigure[Dependency on the headstart $z$ for a fixed $\lambda = 0.2$.]{
            \includegraphics[width=0.45\textwidth]{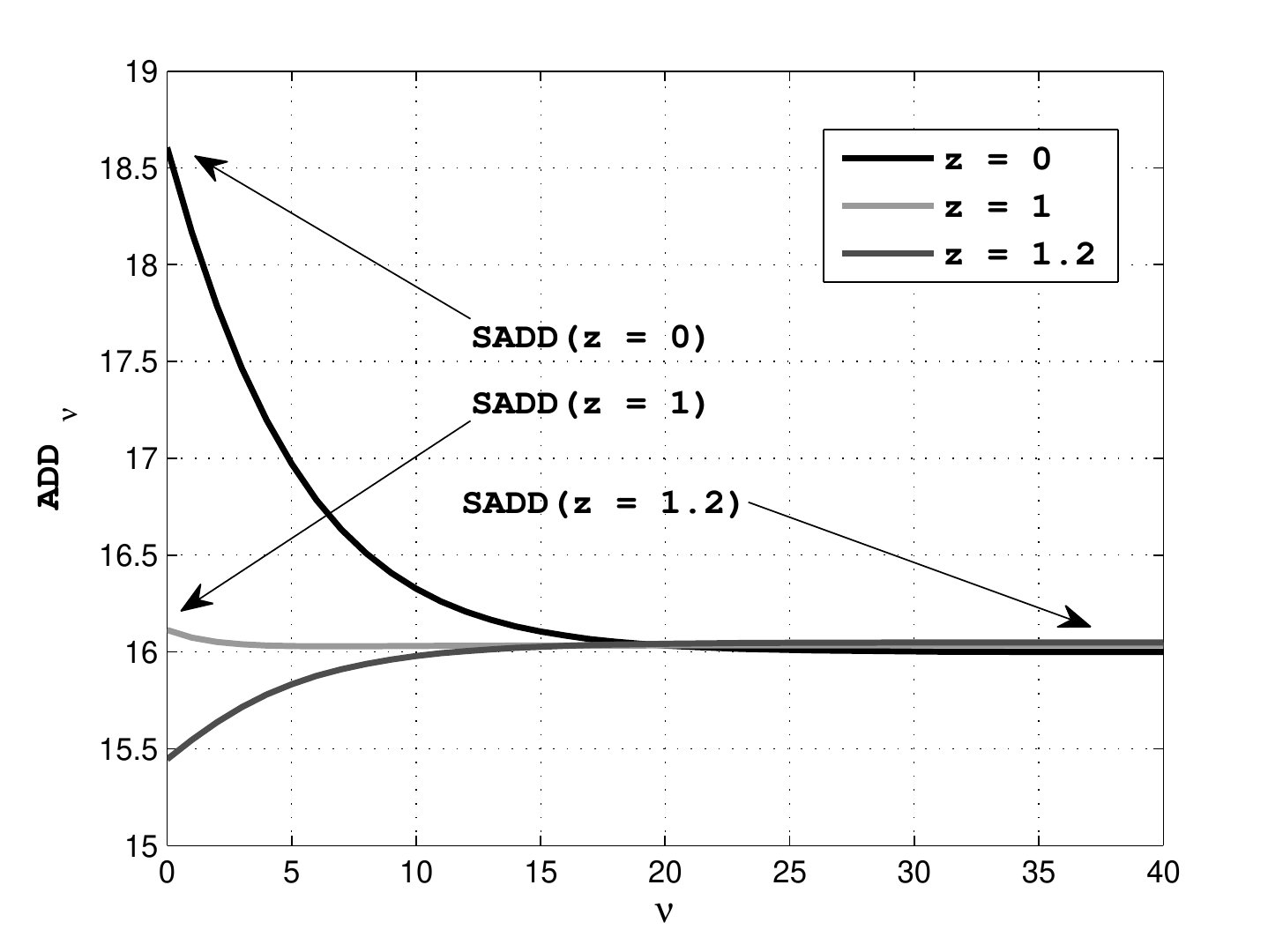}
        }
        \subfigure[Fixed headstart $z = 1$ and $\lambda = 0.1$.]{
            \includegraphics[width=0.45\textwidth]{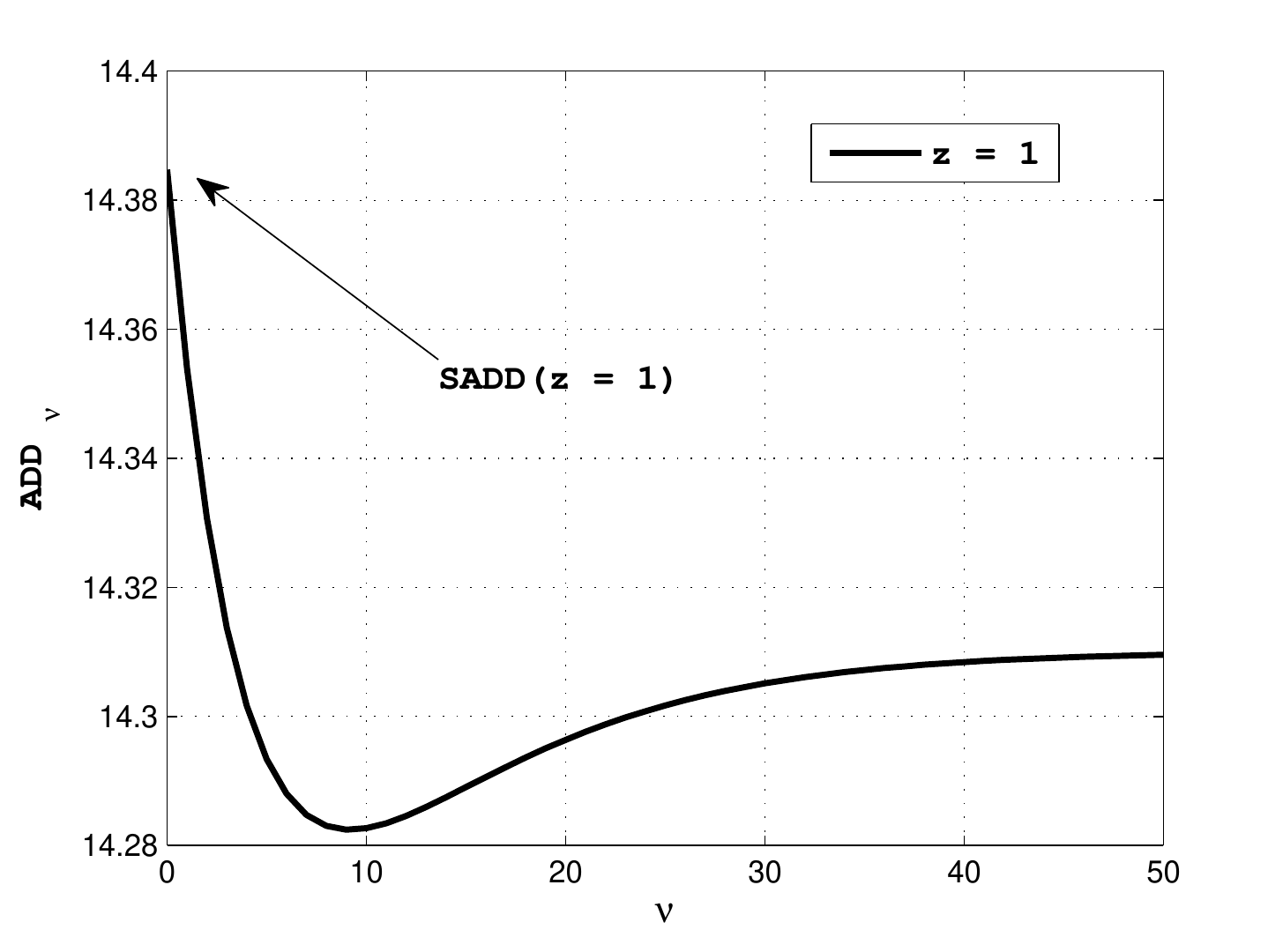}
        }
        \caption{Average detection delay as a function of $\nu $ for a fixed $\ARL = 10^3$.}
        \label{fig:addk_vs_k}
    \end{figure}

   \noindent To get $\set{\delta_k}_{k \ge 1}$ we will look for solutions in the form
    \begin{align}
        \delta _k(z) = B_0^{(k)} + \sum_{n = 1}^\infty B_n^{(k)} \frac{z^n}{n!}, \label{eq:delta_k}
    \end{align}
    subject to $\delta_k(A / \alpha) = 0$.
    The coefficients have the form
    \begin{align*}
        B_0^{(k)} &= e^{-\frac{1}{\lambda}A} \sum_{n = 1}^\infty S_n^{(k - 1)} \frac{A^n}{n!}, \quad
        B_n^{(k)} = \alpha^n \left\{ \frac{1}{\lambda ^n} B_0^{(k)} - S_n^{(k - 1)} \right\},
    \end{align*}
    where
    \[
    S_n^{(k - 1)} = \sum_{j = 1}^n \left(\frac{1}{\lambda}\right)^j B_{n - j}^{(k - 1)}
    \]
    depends on the coefficients from the previous step. When $k = 1$, coefficients from \eqref{eq:delta_zero:coeff} are used.
    Knowing $\set{\rho _k}_{k \ge 1}$ and $\set{\delta_k}_{k \ge 1}$ from \eqref{eq:rho_k} and \eqref{eq:delta_k} respectively, one computes $\ADD _k$ recursively, resulting in $\SADD = \sup_{0 \le k < \infty} \ADD _k$. It can be shown that when the headstart is fixed at $z = 0$ the maximal ADD of EWMA is attained at $\nu=0$, i.e., $\SADD = \ADD_0$. On the other hand, for other initialization points this need not be the case. An example of how $\ADD _\nu $ behaves as a function $\nu $ is shown in Figure~\ref{fig:addk_vs_k}. For each headstart, \eqref{eq:arl} was used to numerically get $\ARL $ at a fixed level.
\vspace{-0.5em}
\subsection{\texorpdfstring{Recovering \boldmath{$\STADD$}}{Recovering stationary ADD}}

  Recall that by \eqref{eq:frac:addk_stadd} $\STADD = {\psi (z)} / {\ell (z)}$,
  where $\ell (z)$ is the $\ARL $ (see \eqref{eq:arl}) and $\psi (z)$ is the integral average detection delay, which satisfies \eqref{eq:int_iadd}:
  \[
    \psi (z) = \delta _0(z) + \frac{1}{\lambda } e^{\frac{1}{\lambda } \alpha z} \int _{\alpha z} ^A e^{-\frac{1}{\lambda }y} \psi (y) \,dy.
  \]
  Here $\delta _0(z)$ is $\ADD _0$, obtained in \eqref{eq:delta_zero}.
  The solution can be written as
  \begin{align*}
      \psi (z) = B_0 \, B_0^{(0)}
	  &- \frac{1}{\lambda } \sum_{n = 2}^\infty \frac{A^n}{n} \frac{[n - 1]_\alpha !}{(n - 1)!} \, s_{n - 1} \\
	  &- \frac{1}{\lambda} \sum_{n = 1}^\infty \frac{(\alpha z)^n}{n} \frac{[n - 1]_\alpha !}{(n - 1)!} \left( \frac{1}{(1 + \theta )^n} + B_0^{(0)} - s_{n - 1} \right),
  \end{align*}
  where $B_0$ and $B_0^{(0)}$ are the coefficients defined in \eqref{eq:arl:coeff} and \eqref{eq:delta_zero:coeff} respectively, and
  \[
      s_0 = 0,  \qquad
      s_n = \sum _{k = 1} ^n \left( \frac{1}{1 + \theta } \right)^k \frac{\alpha ^k}{1 - \alpha ^k}, \quad n \ge 1.
  \]\\
\vspace{-0.5em}

\section{A Case Study}
\label{sec:case-study}

\begin{figure}[!htb]
    \centering
    \subfigure[SADD as a function of $\lambda$ and $z$.]{
        \includegraphics[width=0.45\textwidth]{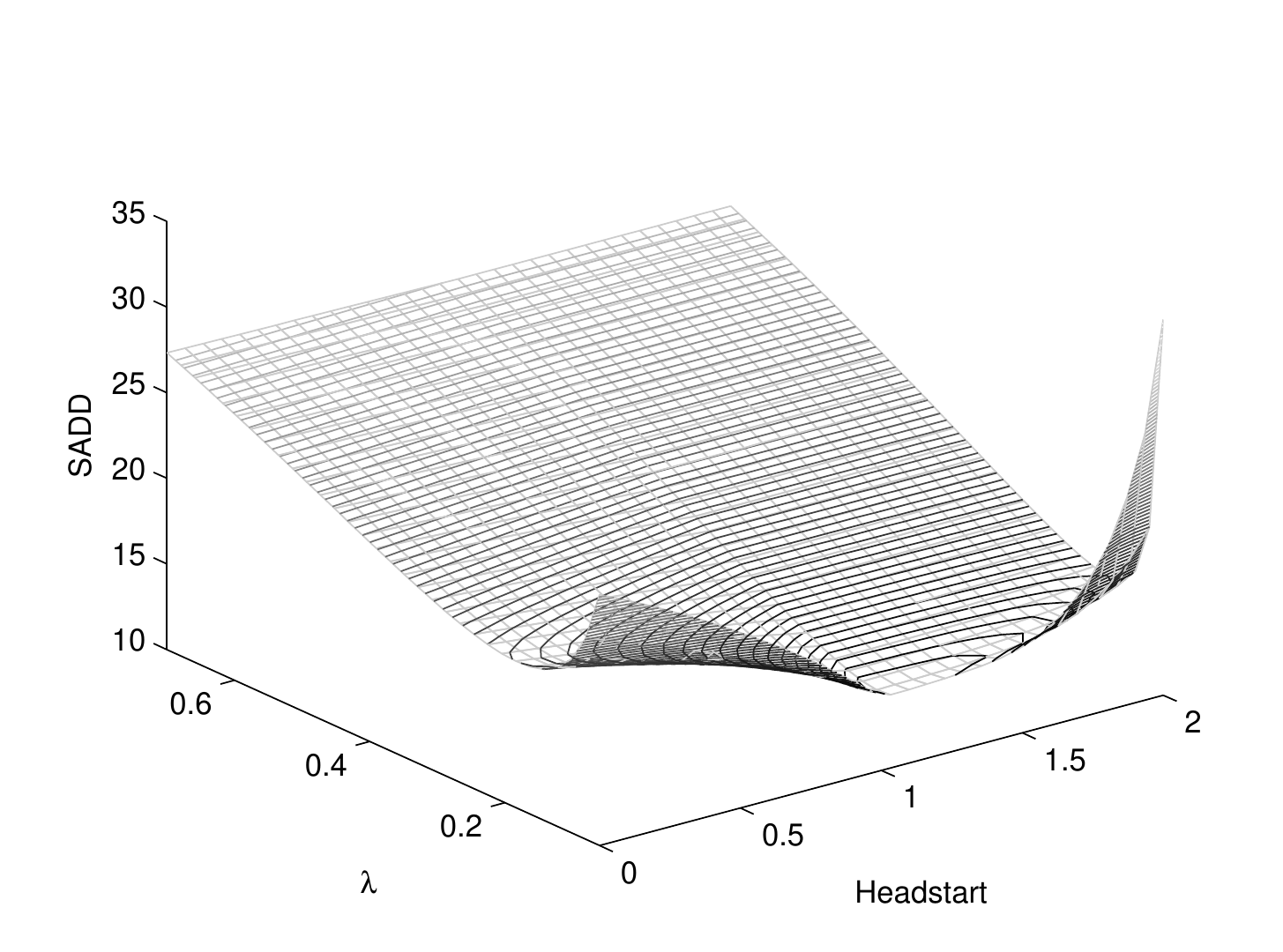}
    }
    \subfigure[STADD as a function of $\lambda$ and $z$.]{
        \includegraphics[width=0.45\textwidth]{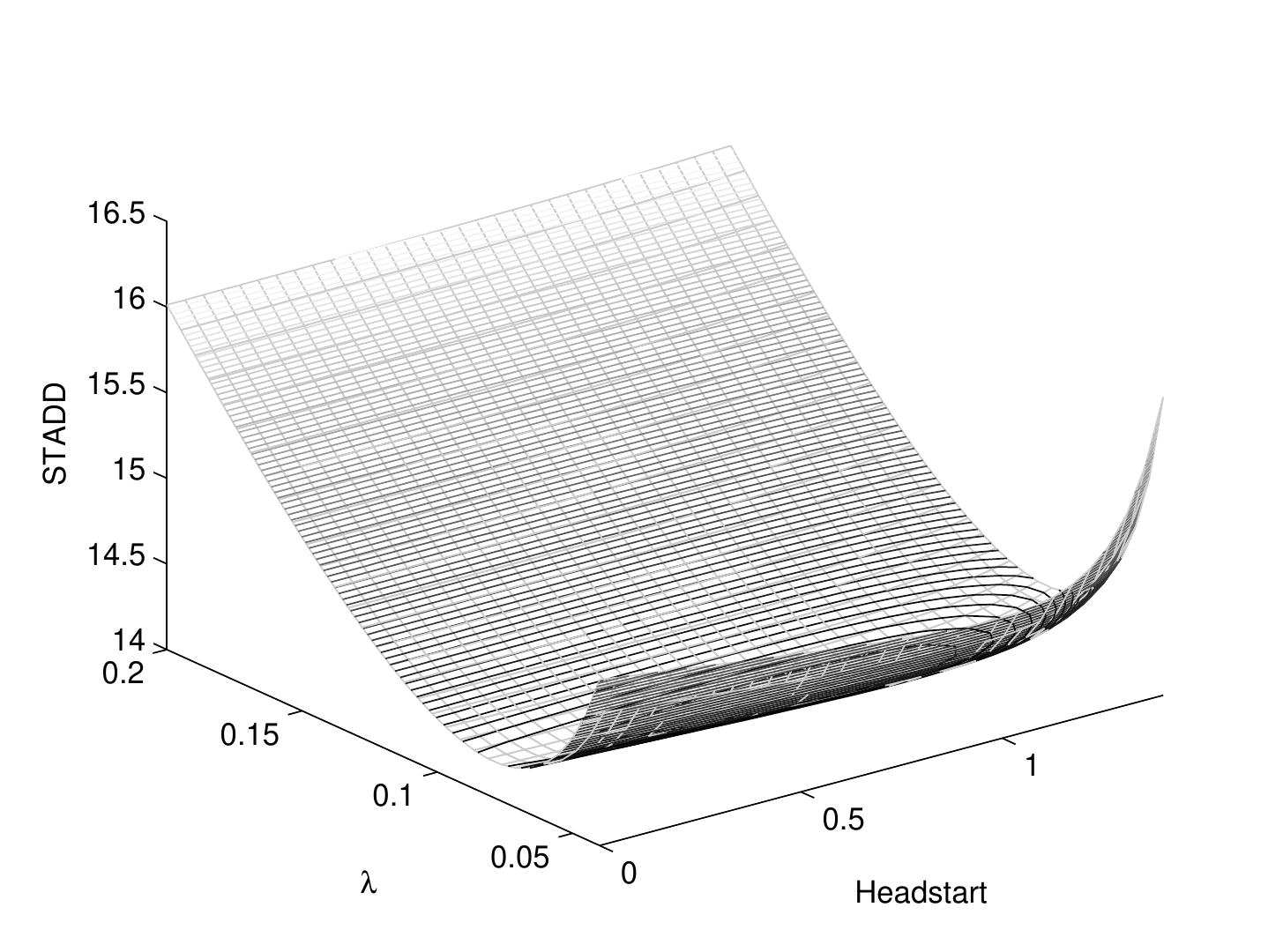}
    }
    \caption{Performance as a function of $\lambda $ and $z$ for a fixed $\ARL = 10^3$.}
    \label{fig:delay_vs_lambda_x_surf}
\end{figure}

We now employ the formulae obtained in the preceding section to compute the performance of the EWMA chart as a function of the smoothing factor $\lambda$ and the headstart $z$. The aim is to find the optimal pair $(\lambda,z)$, i.e., the values of $\lambda$ and $z$ for which the performance of the chart is optimal. Specifically, suppose first that the ARL to false alarm is fixed at a given level, and let us see how EWMA's performance depends on both the smoothing parameter $\lambda $ and the headstart $z$. As can be seen in Figure~\ref{fig:delay_vs_lambda_x_surf}, a proper choice of both $\lambda $ and $z$ may significantly improve the performance of the EWMA procedure.
\noindent We now make two observations based on the numerical results presented in Figure~\ref{fig:delay_vs_lambda_x_surf}. First, regardless of the detection delay measure to be minimized, the impact of the smoothing factor on EWMA's performance is substantial. However, the impact of the headstart, $z$, on EWMA's performance varies depending on the detection delay measure to be minimized. Specifically, one would expect $\SADD$ to be more sensitive to the choice of $z$, and $\STADD$ to be less sensitive. This is in fact confirmed by Figure \ref{fig:delay_vs_lambda_x_surf}.

\begin{figure}[!htb]
    \centering
        \includegraphics[width=0.5\textwidth]{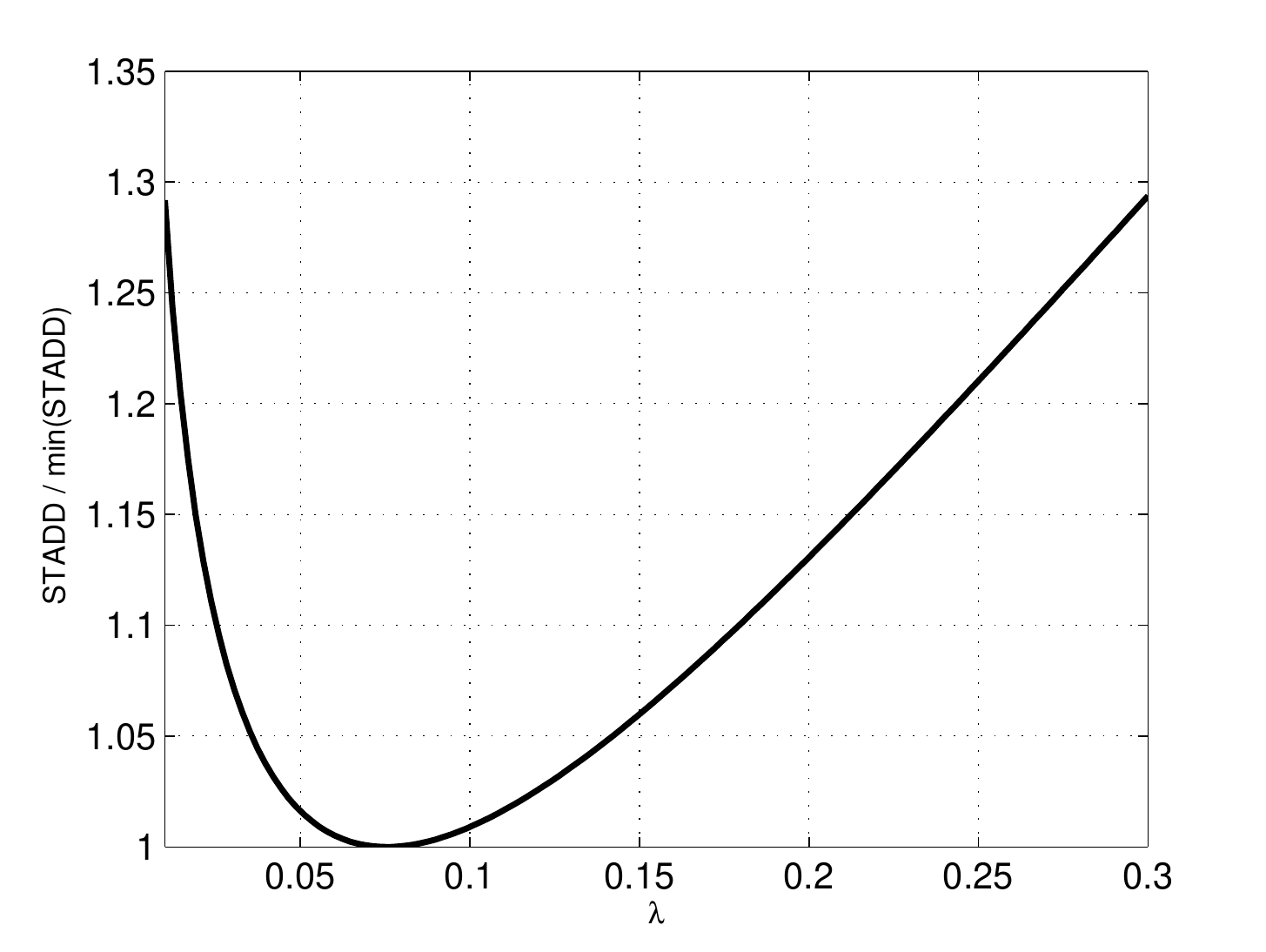}
    \caption{Performance loss as a function of $\lambda $ for fixed $z = 1.0$ and $\ARL = 10^3$.}
    \label{fig:loss_delay_vs_lambda}
\end{figure}

\noindent Comparison results of EWMA optimized in $\lambda$, but not in $z$, to optimal procedures are summarized in Tables \ref{tbl:compare_mean_15} and \ref{tbl:compare_mean_20}. Recall that SR--$r$ is asymptotically third-order minimax and SR is exactly optimal in the stationary setting. Tables show that optimized EWMA has almost the same performance as SR--$r$ in the minimax setting and as SR w.r.t. $\STADD$. Moreover, evidence suggests that for the exponential scenario fixing the headstart at $z = 1$ does not lead to significant drop in performance. On the other hand, optimizing over $\lambda $ is crucial. How the performance is affected by $\lambda$ w.r.t.\ $\SADD$ when the headstart is fixed at $z = 1$ is shown in Figure \ref{fig:loss_delay_vs_lambda}.

\begin{table}[!ht]
    \centering
     \caption{Shift in mean $\theta = 0.5$. In each case $\lambda $ is chosen optimally for given $z$ and $\gamma$.}
    \label{tbl:compare_mean_15}
    \subtable[Supremum average detection delay.]{
        \begin{tabular}{llccc}
        \hline  & $\gamma$ & $10^2$  & $10^3$  & $10^4$ \\
        \hline EWMA
            & $\SADD$   & $17.7$  & $46.5$  & $85.5$  \\
            $z = 0$
            & $A$       & $2.07$  & $1.79$  & $1.67$  \\
            & $\lambda$ & $0.275$ & $0.096$ & $0.049$ \\
        \hline EWMA
            & $\SADD$   & $14.8$  & $33.4$  & $58.1$  \\
            $z = 1$
            & $A$       & $1.39$  & $1.37$  & $1.38$  \\
            & $\lambda$ & $0.086$ & $0.035$ & $0.021$ \\
        \hline EWMA
            & $\SADD$   & $14.7$  & $33.4$  & $58.1$  \\
            $z = z_{opt}$
            & $A$       & $1.30$  & $1.37$  & $1.38$  \\
            & $\lambda$ & $0.067$ & $0.035$ & $0.021$ \\
            & $z$       & $0.96$  & $1.04$  & $1.00$  \\
        \hline SR-$r$
            & $\SADD$   & $14.7$  & $33.3$  & $56.4$  \\
            \hline
        \end{tabular}
    }
    \subtable[Stationary average detection delay.]{
        \begin{tabular}{llccc}
        \hline  & $\gamma$ & $10^2$  & $10^3$  & $10^4$ \\
        \hline EWMA
            & $\STADD$  & $14.4$  & $33.6$  & $57.6$  \\
            $z = 0$
            & $A$       & $1.38$  & $1.41$  & $1.38$  \\
            & $\lambda$ & $0.095$ & $0.040$ & $0.021$ \\
        \hline EWMA
            & $\STADD$  & $14.7$  & $33.4$  & $57.5$  \\
            $z = 1$
            & $A$       & $1.35$  & $1.37$  & $1.37$  \\
            & $\lambda$ & $0.077$ & $0.035$ & $0.021$ \\
        \hline EWMA
            & $\STADD$  & $14.3$  & $33.3$  & $57.5$  \\
            $z = z_{opt}$
            & $A$       & $1.31$  & $1.37$  & $1.37$  \\
            & $\lambda$ & $0.076$ & $0.035$ & $0.021$ \\
            & $z$       & $0.53$  & $0.84$  & $0.93$  \\
        \hline SR
            & $\STADD$  & $14.3$  & $32.8$  & $55.6$  \\
            \hline
        \end{tabular}
    }
\end{table}

\begin{table}[!ht]
    \centering
     \caption{Shift in mean $\theta = 1.0$. In each case $\lambda $ is chosen optimally for given $z$ and $\gamma$.}
    \label{tbl:compare_mean_20}
    \subtable[Supremum average detection delay.]{
        \begin{tabular}{llccc}
        \hline  & $\gamma$ & $10^2$  & $10^3$  & $10^4$ \\
        \hline EWMA
            & $\SADD$   & $8.99$  & $18.6$  & $30.1$  \\
            $z = 0$
            & $A$       & $2.55$  & $2.29$  & $2.13$  \\
            & $\lambda$ & $0.412$ & $0.181$ & $0.102$ \\
        \hline EWMA
            & $\SADD$   & $7.56$  & $14.2$  & $22.1$  \\
            $z = 1$
            & $A$       & $1.61$  & $1.64$  & $1.67$  \\
            & $\lambda$ & $0.142$ & $0.073$ & $0.049$ \\
        \hline EWMA
            & $\SADD$   & $7.54$  & $14.2$  & $22.1$  \\
            $z = z_{opt}$
            & $A$       & $1.58$  & $1.72$  & $1.67$  \\
            & $\lambda$ & $0.134$ & $0.085$ & $0.049$ \\
            & $z$       & $0.96$  & $1.13$  & $1.00$  \\
        \hline SR-$r$
            & $\SADD$   & $7.5$   & $14.2$  & $21.5$ \\
            \hline
        \end{tabular}
    }
    \subtable[Stationary average detection delay.]{
        \begin{tabular}{llccc}
        \hline  & $\gamma$ & $10^2$  & $10^3$  & $10^4$ \\
        \hline EWMA
            & $\STADD$  & $7.51$  & $14.2$  & $22.0$  \\
            $z = 0$
            & $A$       & $1.64$  & $1.68$  & $1.67$  \\
            & $\lambda$ & $0.156$ & $0.079$ & $0.049$ \\
        \hline EWMA
            & $\STADD$  & $7.54$  & $14.2$  & $21.9$  \\
            $z = 1$
            & $A$       & $1.58$  & $1.66$  & $1.67$  \\
            & $\lambda$ & $0.136$ & $0.075$ & $0.049$ \\
        \hline EWMA
            & $\STADD$  & $7.49$  & $14.2$  & $21.9$  \\
            $z = z_{opt}$
            & $A$       & $1.58$  & $1.66$  & $1.67$  \\
            & $\lambda$ & $0.138$ & $0.076$ & $0.049$ \\
            & $z$       & $0.54$  & $0.82$  & $0.92$  \\
        \hline SR
            & $\STADD$  & $7.45$  & $13.9$  & $21.2$  \\
            \hline
        \end{tabular}
    }
\end{table}
\newpage

\begin{figure}[!htb]
    \centering
    \subfigure[$\STADD$ as a function of $\lambda$ and ARL.]{
        \includegraphics[width=0.45\textwidth]{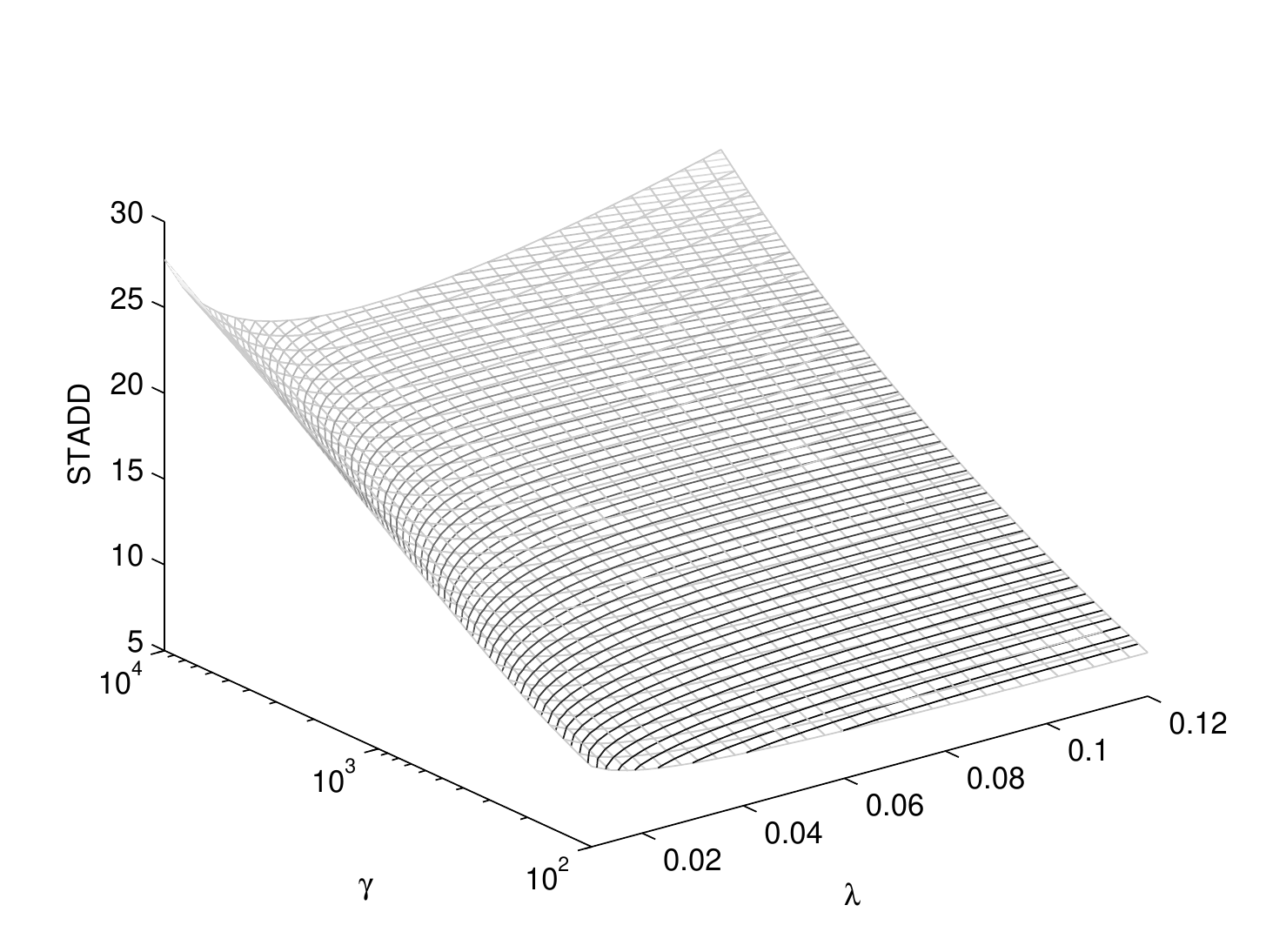}
    }
    \subfigure[Optimal choice of the smoothing parameter $\lambda $.]{
        \includegraphics[width=0.45\textwidth]{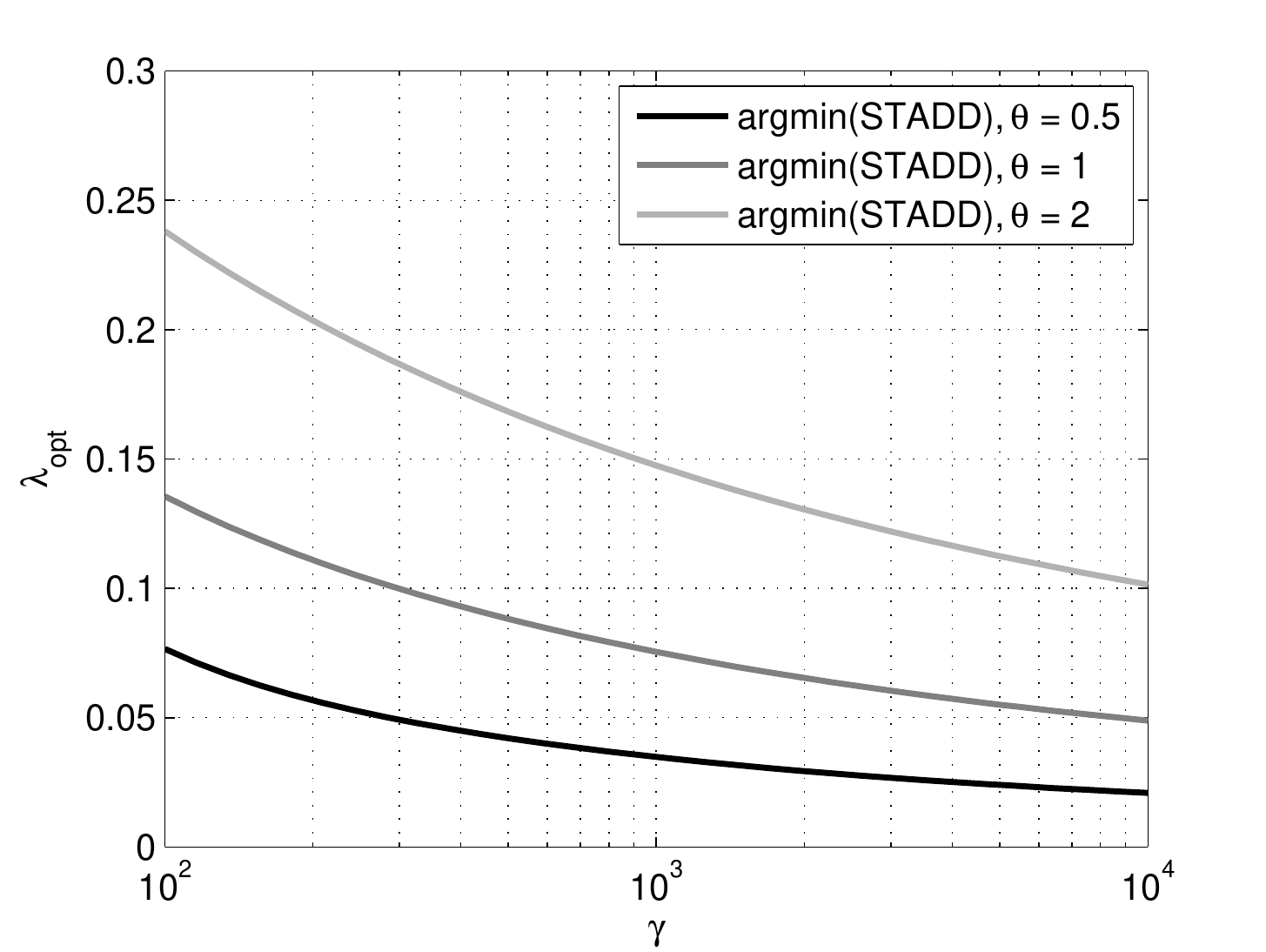}
    }
    \caption{Optimal choice of $\lambda $ as a function of $\ARL = \gamma $; the headstart is fixed at $z = 1$.}
    \label{fig:delay_vs_arl}
\end{figure}

\noindent Next, we consider the performance as a function of $\ARL$. Since we have two other parameters to optimize over, we will present the case where the headstart is fixed at $z = 1$ and optimize $\STADD$ w.r.t. $\lambda$. Figure~\ref{fig:delay_vs_arl} suggests that $\lambda_{opt} \rightarrow 0$ as $\gamma \rightarrow \infty$.

\begin{figure}[!htb]
    \centering
    \subfigure[Performance loss ratio for EWMA.]{
        \includegraphics[width=0.45\textwidth]{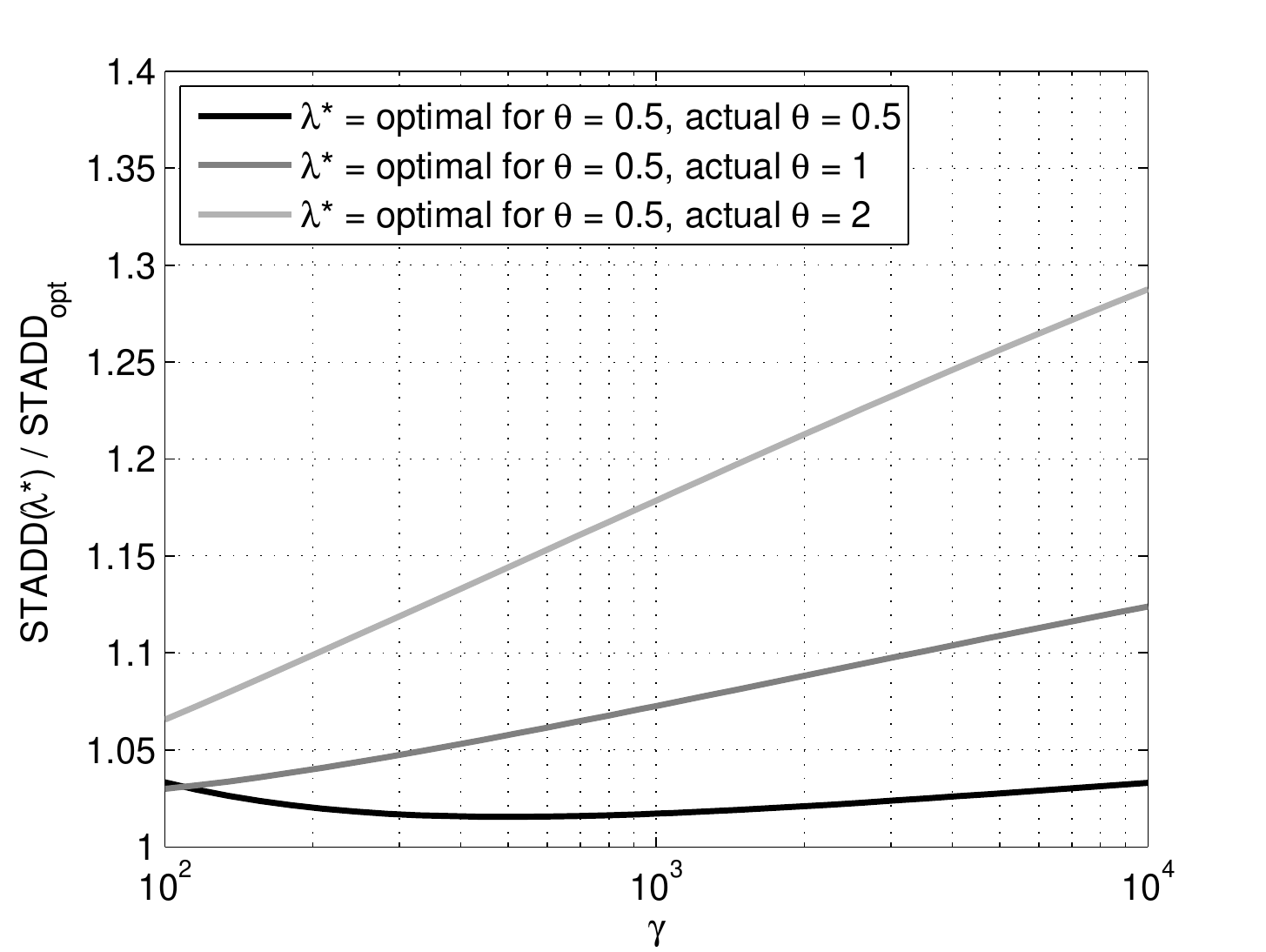}
    }
    \subfigure[Performance loss ratio for SR.]{
        \includegraphics[width=0.45\textwidth]{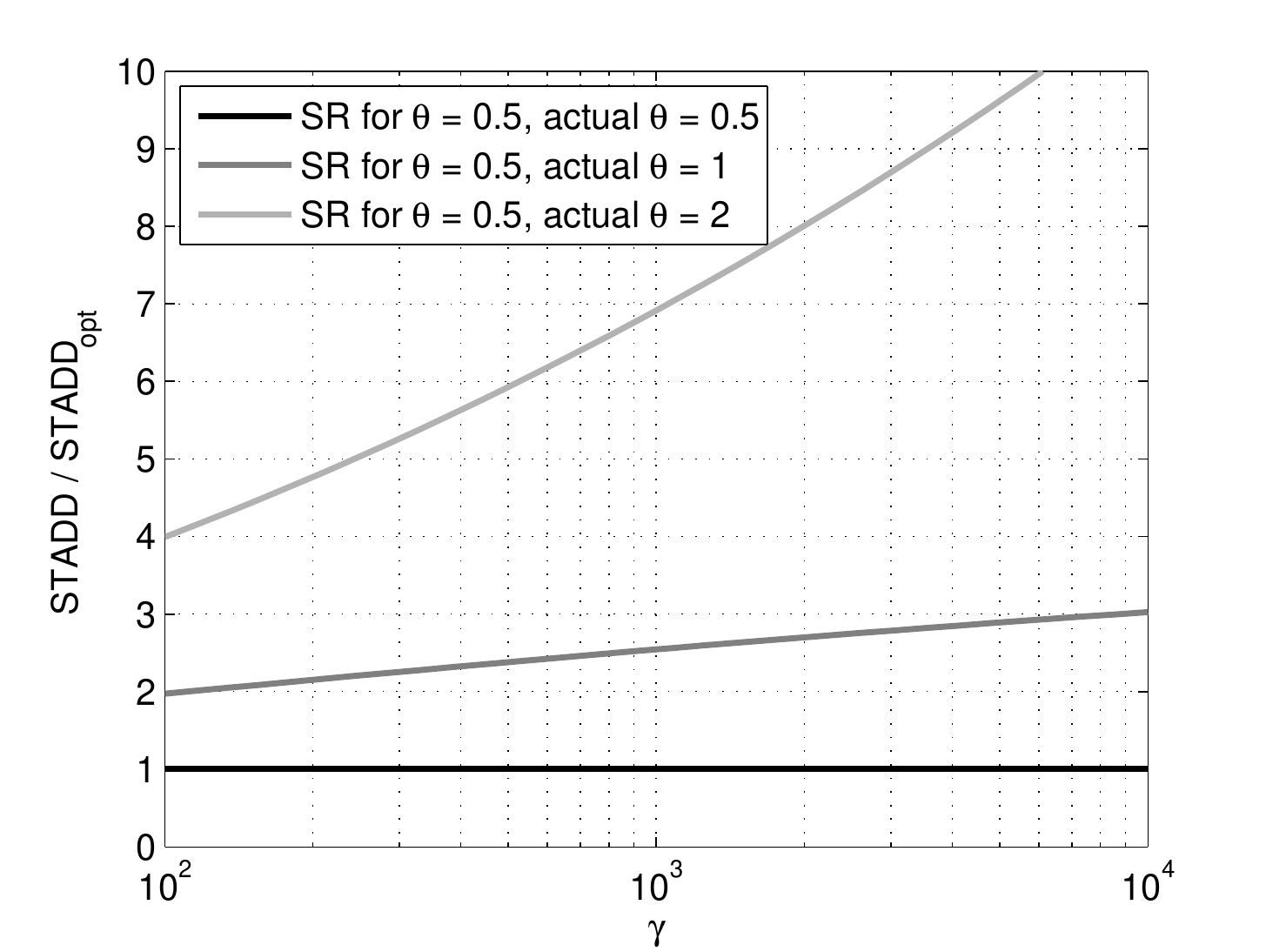}
    }
\caption{The performance loss when tuned to a wrong post-change parameter; the headstart is fixed at $z = 1$.}
\label{fig:discord}
\end{figure}

\noindent Lastly, we consider a case where the post-change parameter is unknown, but lies within a certain range, say $\theta \ge 0.5$. Choosing an optimal $\lambda$ for the worst-case scenario, namely for the smallest change, one might be interested in how it affects performance should the post-change parameter be different. For calculations we take $\theta = 0.5, 1.0, 2.0$. As a benchmark, we take the SR procedure tuned to the correct $\theta $, which is exactly optimal in the stationary sense and inspect the loss in performance for \begin{inparaenum}[\itshape a\upshape)] \item EWMA with $\lambda $ chosen optimally for $\theta = 0.5$, and \item SR procedure designed with $\theta = 0.5$ in mind\end{inparaenum}. The results are summarized in Figure~\ref{fig:discord}. It can be seen that in such a setup EWMA has an upper hand. Indeed, it is less sensitive to the parameter misspecification.\\
\vspace{-0.5em}

\section*{Acknowledgements}

The work of A.\ G.\ Tartakovsky was supported in part by the U.S.\ Air Force Office of Scientific Research under MURI grant FA9550-10-1-0569, by the U.S.\ Defense Threat Reduction Agency under grant HDTRA1-10-1-0086, by the U.S.\  Defense Advanced Research Projects Agency under grant W911NF-12-1-0034, by the U.S.\ National Science Foundation under grant DMS-1221888, and by the U.S.\ Army Research Office under grants W911NF-13-1-0073 and W911NF-14-1-0246  at the University of Southern California, Department of Mathematics and at the University of Connecticut, Department of Statistics.

\noindent A.\ S.\ Polunchenko is personally thankful to Prof.\ Sven Knoth of the Helmut Schmidt University, Hamburg, Germany, for providing constructive feedback on an earlier version of the paper and for pointing out many important and relevant references.

\noindent Finally, we would like to thank Prof.\ Nitis Mukhopadhyay for the invitation to publish this work and for the time and effort he spent to produce this special issue.

\newpage

\setcitestyle{numbers}
\renewcommand\bibnumfmt[1]{#1.}

\vfill\vfill

\end{document}